\documentclass[a4paper,nodate]{sab} 

\usepackage{graphicx}
\usepackage[backend=biber, 
            natbib=true, 
            sortcites, 
            sorting=nyt, 
            doi=false, 
            url=false, 
            eprint=false, 
            style=apa
            ]{biblatex}
\usepackage{csquotes}
\usepackage{txfonts} 
\usepackage{hyperref}
\usepackage{url}
\usepackage[T1]{fontenc} 
\usepackage{t1enc}
\usepackage{xcolor}

\newcommand{\feh}{[Fe/H]}
\newcommand{\mgfe}{[Mg/Fe]}
\newcommand{\teff}{$T_{\rm eff}$}
\newcommand{\logg}{$\log g$}

\renewcommand\ackname{Acknowledgements.} 
\renewcommand\andname{\&}
\renewcommand\keywordname{Keywords.} 
\renewcommand\figureptname{Figure}
\renewcommand\tableptname{Table}

\renewcommand\noteaddname{Note added to proofs.}
\renewcommand\andname{\&}

\idline{Boletim da Sociedade Astron\^omica Brasileira, {\bf 31}, no. 1, XX-XX}{1}

\begin{document} 

\title{Stellar properties and chemical features of the \textit{Gaia} Catalogue of Nearby Stars observed by GALAH DR4} 

\author{P. H. R. de Andrade\inst{1} 
        \and 
        M. L. L. Dantas \inst{2,3} 
        \and
        A. C. Soja
        \inst{1} 
        } 

\institute{
    Instituto Federal de Educação, Ciência e Tecnologia Fluminense \emph{Campus} Bom Jesus do Itabapoana, Rio de Janeiro, Brazil. \\
    \email{pedroiff0@gmail.com}  
    \and 
    Instituto de Astrofísica, Pontifícia Universidad Católica de Chile, Vicuña Mackenna 4860, Santiago, Chile. 
    \and 
    Centro de Astro-Ingeniería, Pontifícia Universidad Católica de Chile, Vicuña Mackenna 4860, Santiago, Chile.
}
\date{Received} 

\Abstract{
The \emph{Gaia} Catalogue of Nearby Stars (GCNS) comprises approximately 330\,000 stars within 100 pc of the Sun, as observed by \emph{Gaia} data release 3 (\emph{Gaia} DR3). Meanwhile, the GALAH DR4 survey has spectroscopically characterised nearly one million stars, delivering detailed chemical abundances (up to 30 elements). We present a joint analysis of the $\sim 6\,000$ stars common to both catalogues, offering initial insights into the stellar and chemical properties of the solar neighborhood. Our preliminary results indicate that the majority of these stars are FGK main-sequence objects, with some A-type interlopers (with effective temperatures ranging between 3\,000 and 8\,000\,K), with median ages of $\sim$ 1.6 Gyr (ranging from 0.10 to 14.79 Gyr), and metal-poorer when compared to the Sun: \feh\ $\approx$ -0.19 dex. Additionally, most of the stars are disc members, with some local halo (high-velocity) stars identified. Building on this foundation, future work will deeper exploit the full spectroscopic information and orbital parameters from value-added catalogues to refine Galactic component classifications (thin-thick disc versus halo membership), perform detailed chemical profiling, and deliver a comprehensive chemo-dynamical characterisation of the solar neighborhood. This will provide new insights into the formation and evolution of nearby stellar populations.}
{O \emph{Gaia Catalogue of Nearby Stars} do \emph{Gaia} (GCNS) compreende aproximadamente 330\,000 estrelas situadas até 100\,pc do Sol, conforme observado pelo \emph{Gaia} DR3. Por sua vez, o levantamento GALAH DR4 caracterizou espectroscopicamente cerca de um milhão de estrelas, fornecendo dados de alto valor que incluem abundâncias químicas detalhadas de até 30 elementos. Apresentamos uma análise conjunta das cerca de 6\,000 estrelas em comum entre ambos os catálogos, oferecendo uma visão inicial sobre as propriedades estelares e químicas da vizinhança solar. Nossos resultados preliminares indicam que a maioria dessas estrelas são objetos de sequência principal do tipo FGK, com algumas de tipo A (com temperaturas efetivas variando entre 3\,000 e 8\,000\,K), apresentando idades medianas de $\sim$ 1.6\,Ga (no intervalo de 0.10 a 14.79\,Ga) e metalicidade inferior à do Sol: \feh\ $\approx$ -0.19 dex. Ademais, a maioria das estrelas são membros do disco, com algumas identificadas como pertencentes ao halo (alta velocidade). Com base nesses resultados, trabalhos futuros explorarão mais profundamente as informações espectroscópicas completas e os parâmetros orbitais provenientes de catálogos de valor agregado, a fim de refinar as classificações dos componentes Galácticos (disco espesso-fino versus halo), realizar perfis químicos detalhados e fornecer uma caracterização quimiodinâmica abrangente da vizinhança solar. Isso trará novas perspectivas sobre a formação e a evolução das populações estelares próximas.
}


\titlerunning{Stellar properties and chemical features of the Gaia Catalogue of Nearby Stars observed by GALAH DR4}

\authorrunning{Andrade et al.}

\maketitle 

\section{Introduction}
\label{sec:intro}
        
The detailed characterization of stars in the solar neighborhood offers a high-resolution view of the Milky Way’s (MW) formation and evolution. Variations in elemental abundances and orbital motions act as chemo-dynamical fingerprints, tracing stellar origins, co-eval groups, and the enrichment processes that shaped the Solar System’s local  environment. In this context, \emph{Gaia} mission \citep[][]{Gaia2016} has delivered precise astrometric data for nearby MW stars, notably through the \emph{Gaia} Catalogue of Nearby Stars \citep[GCNS;][comprising $\sim$ 330\,000 stars within 100\,pc, derived from the third \emph{Gaia} data release, DR3]{GaiaGCNS}.

Complementary to \emph{Gaia}, large-scale spectroscopic surveys provide the stellar parameters and chemical abundances needed to investigate the chemo-dynamical structure of the Milky Way. Notable examples include \emph{Gaia}-ESO \citep[][]{Gilmore2022, Randich2022}, LAMOST \citep[][]{Cui2012, Xiang2019}, and APOGEE \citep[][]{Majewski2017}. Among these, the Galactic Archeology with HERMES survey \citep[GALAH;][]{Buder2025} has observed nearly one million stars in his fourth data release (DR4), delivering up to 30 elemental abundances and value-added catalogs with stellar kinematics.

In this work, we perform a joint analysis of $\sim$6\,000 stars cross-matched between the GCNS and GALAH~DR4. By combining astrometric and spectroscopic information, we aim to characterize in detail the stellar populations within 100~pc, focusing on their ages and chemical abundances.

\vspace{-0.3cm}

\section{Data and methodology}
\label{sec:data_and_methods}

We combined the astrometric data from the GCNS \citep[][]{GaiaGCNS} with the chemo-dynamic parameters from GALAH DR4 \citep[][]{Buder2025} and their associated added-value catalogues (which include dynamic parameters estimated via \textsc{Galpy}; \citealt{Bovy2015}, using the \citealt{McMillan2017} potential as input), yielding a sample of over 6\,000 stars. In GALAH DR4, age determination was performed using isochrone fitting from PARSEC+COLIBRI tracks \citep[][]{Bressan2012, Marigo2017}. Spurious data in terms of \feh, age, \teff, \mgfe, \logg\ were also excluded from the final dataset. 

\vspace{-0.3cm}

\section{Analysis, discussion, and final remarks}
\label{sec:analysis}

\begin{figure}[!ht]
    \centering
    \includegraphics[width=0.49\linewidth]{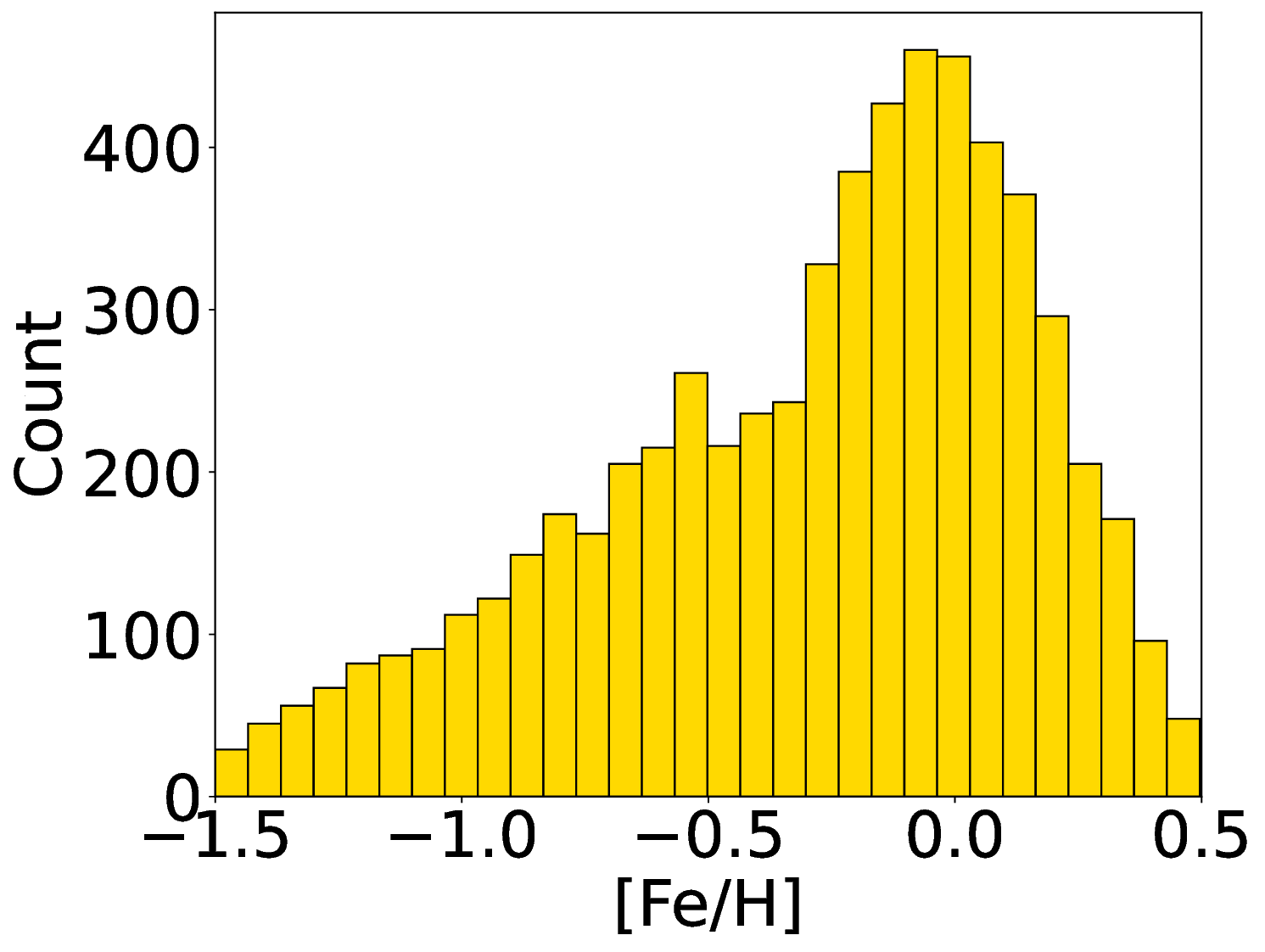}
    \includegraphics[width=0.49\linewidth]{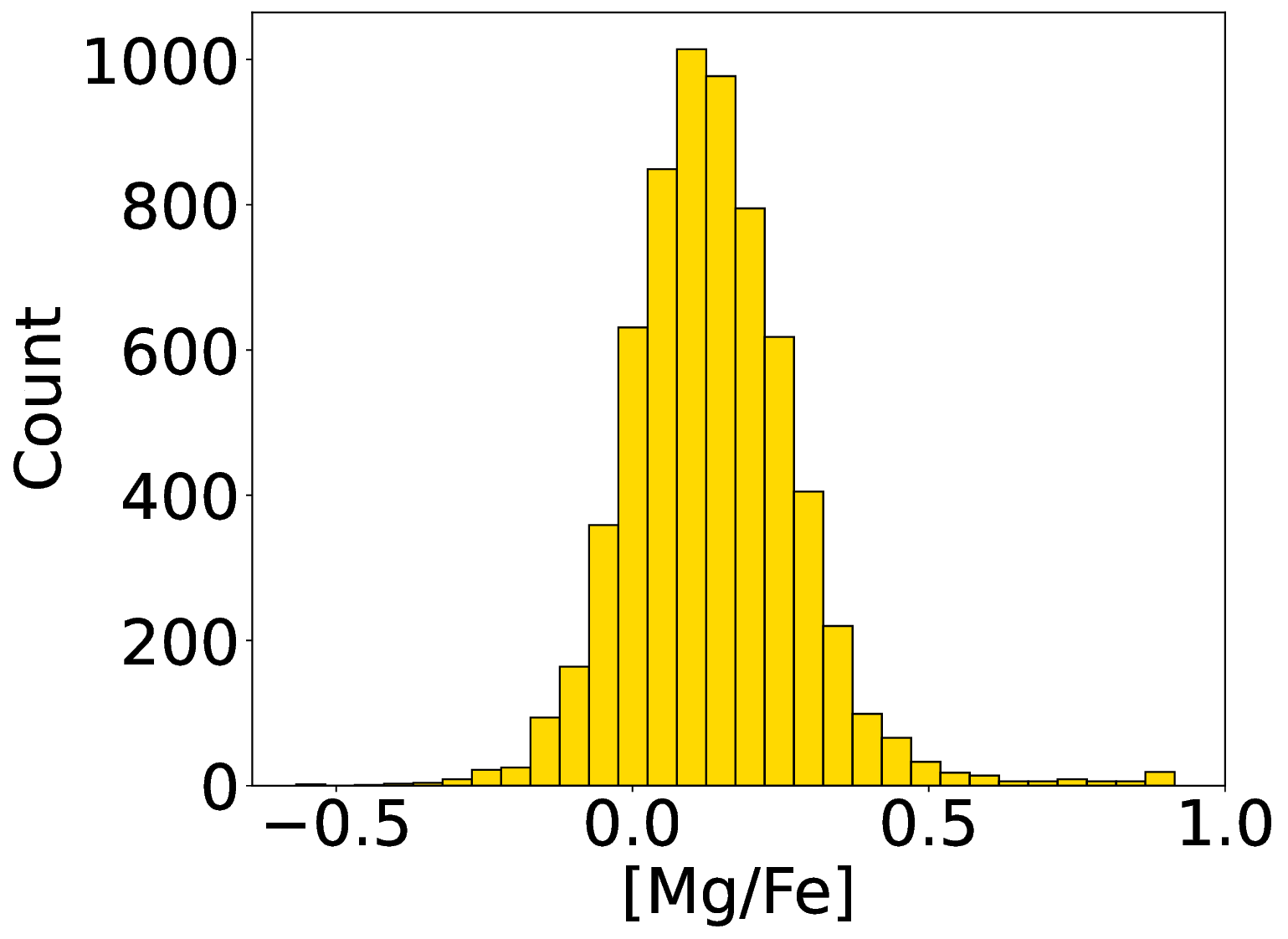}
    \caption{\small \feh\ (right) and \mgfe\ (left) distributions of our sample.}
    \label{fig:histograms}
\end{figure}

The preliminary joint analysis of the approximately 6\,000 stars in the solar neighbourhood matched between GCNS and GALAH DR4 reveals a population predominantly composed of FGK main-sequence stars and some A-type interlopers, with a median age of $\sim$ 1.6 Gyr and slightly more metal-poor when compared to the Sun (median \feh\ $\approx$ -0.19 dex). The parameters examined included chemical features, such as iron \feh\ and \mgfe\ as shown in Fig. \ref{fig:histograms}.  

\begin{figure}[!ht]
    \centering
    \includegraphics[width=0.49\linewidth]{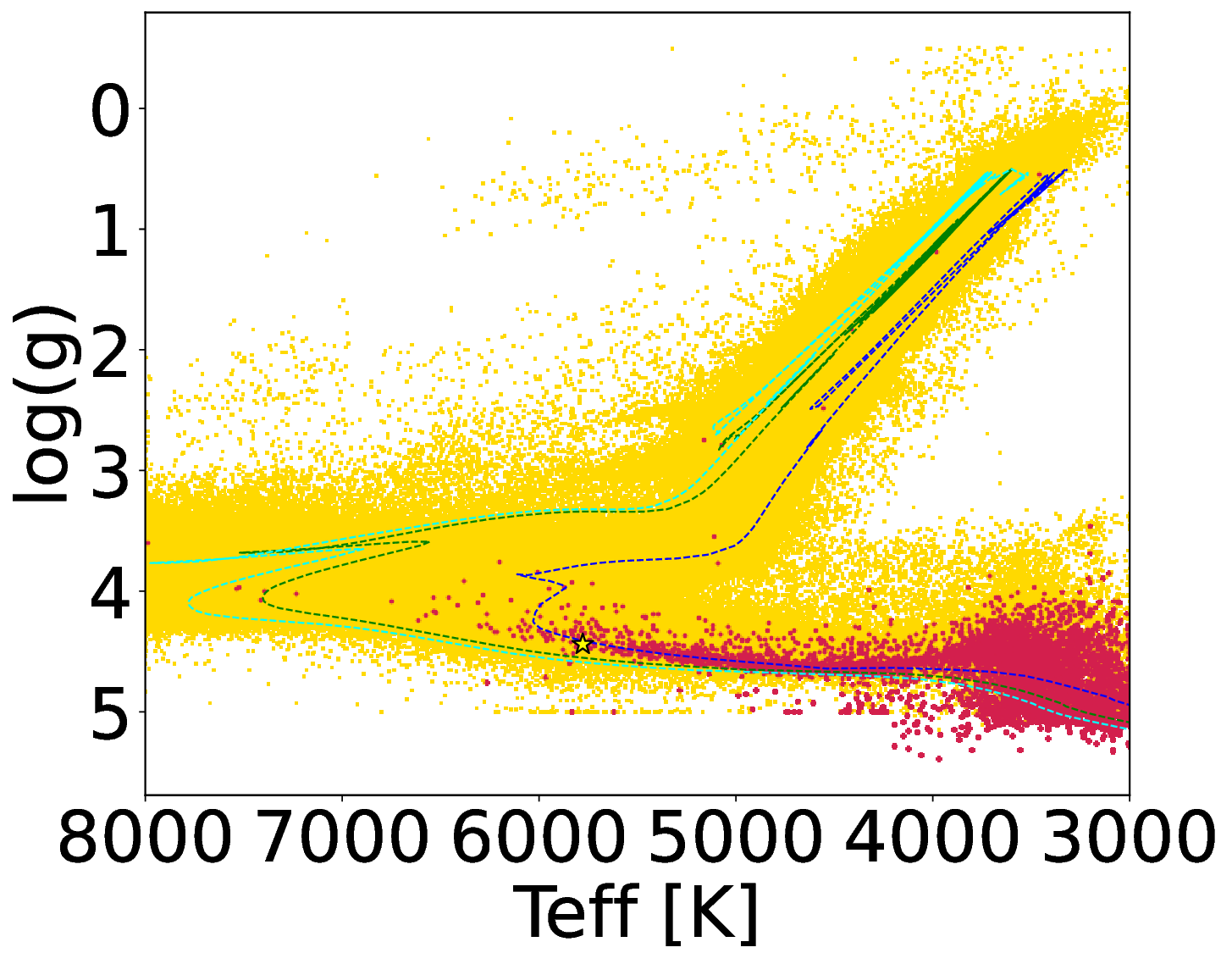}
    \includegraphics[width=0.49\linewidth]{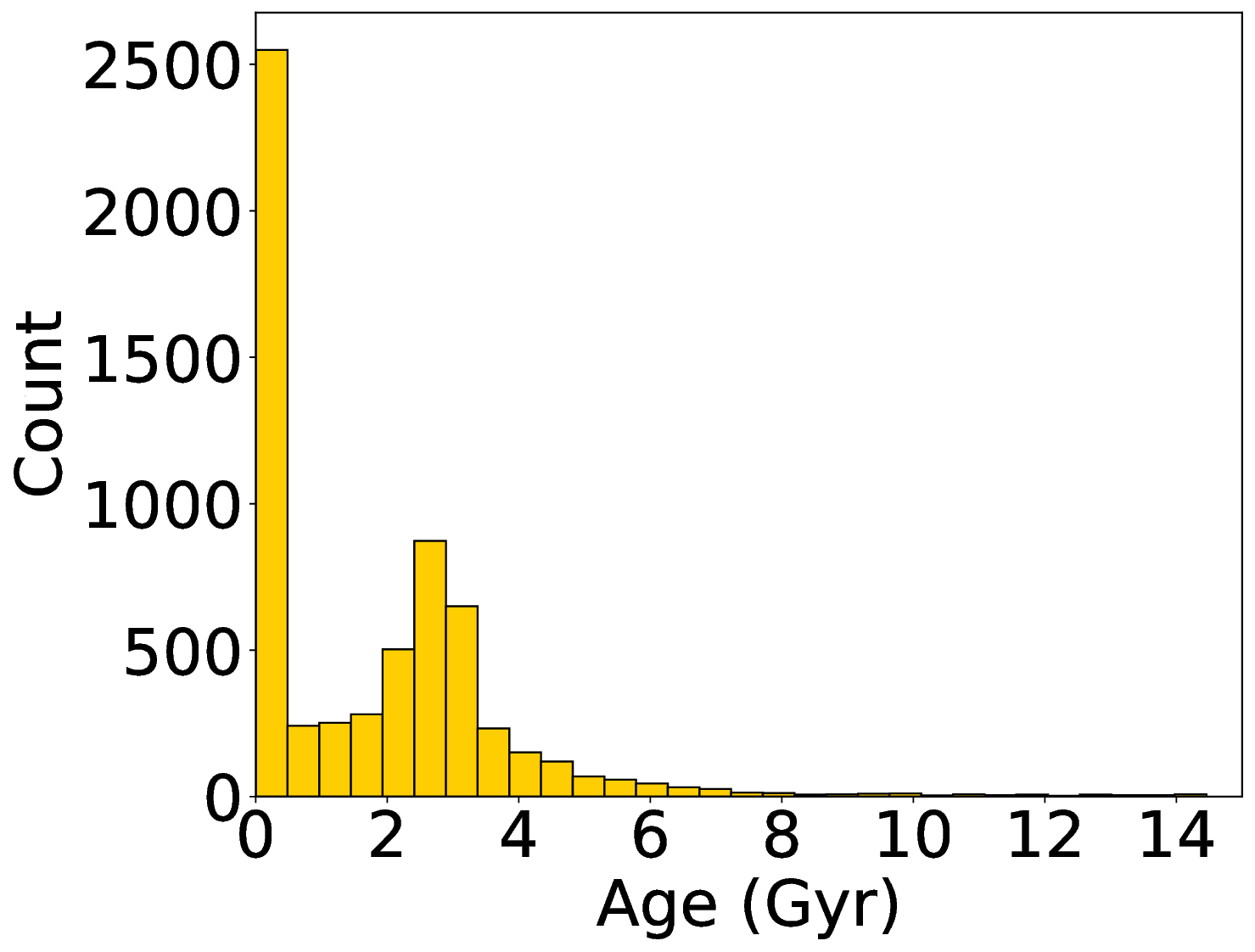}
    \caption{\small Kiel diagram and age distribution of our sample. Left panel: Kiel diagram with PARSEC isochrones overlaid \citep{Bressan2012} centred at $\sim$ 1.6 Gyr and \feh\ $\approx -0.19$ dex (median values of our sample; green), and $\pm1 \sigma$ (cyan and blue) at fixed median \feh. For reference, the full GALAH DR4 is shown in yellow, while our targets are in red. Right panel: stellar age distribution of our sample.}
    \label{fig:kiel}
\end{figure}

The Kiel Diagram (Fig. \ref{fig:kiel}, left panel) depicts the combined sample of GCNS and GALAH DR4 (in red) and the entire GALAH DR4 (in yellow), with some PARSEC \citep{Bressan2012} isochrones overlaid for reference (median in green, $\pm 1 \sigma$ in blue and cyan). The age distribution (Fig. \ref{fig:kiel}, right panel) reveals an apparent excess of young stars; however, it remains uncertain whether these objects are genuinely young or simply occupy a region of the diagram where isochrones are essentially degenerate. This may represent either a real artifact or, more likely, a bias arising from the stars' location within a parameter space characterised by extremely large age uncertainties.

\begin{figure}[!ht]
    \centering
    \includegraphics[width=0.6\linewidth]{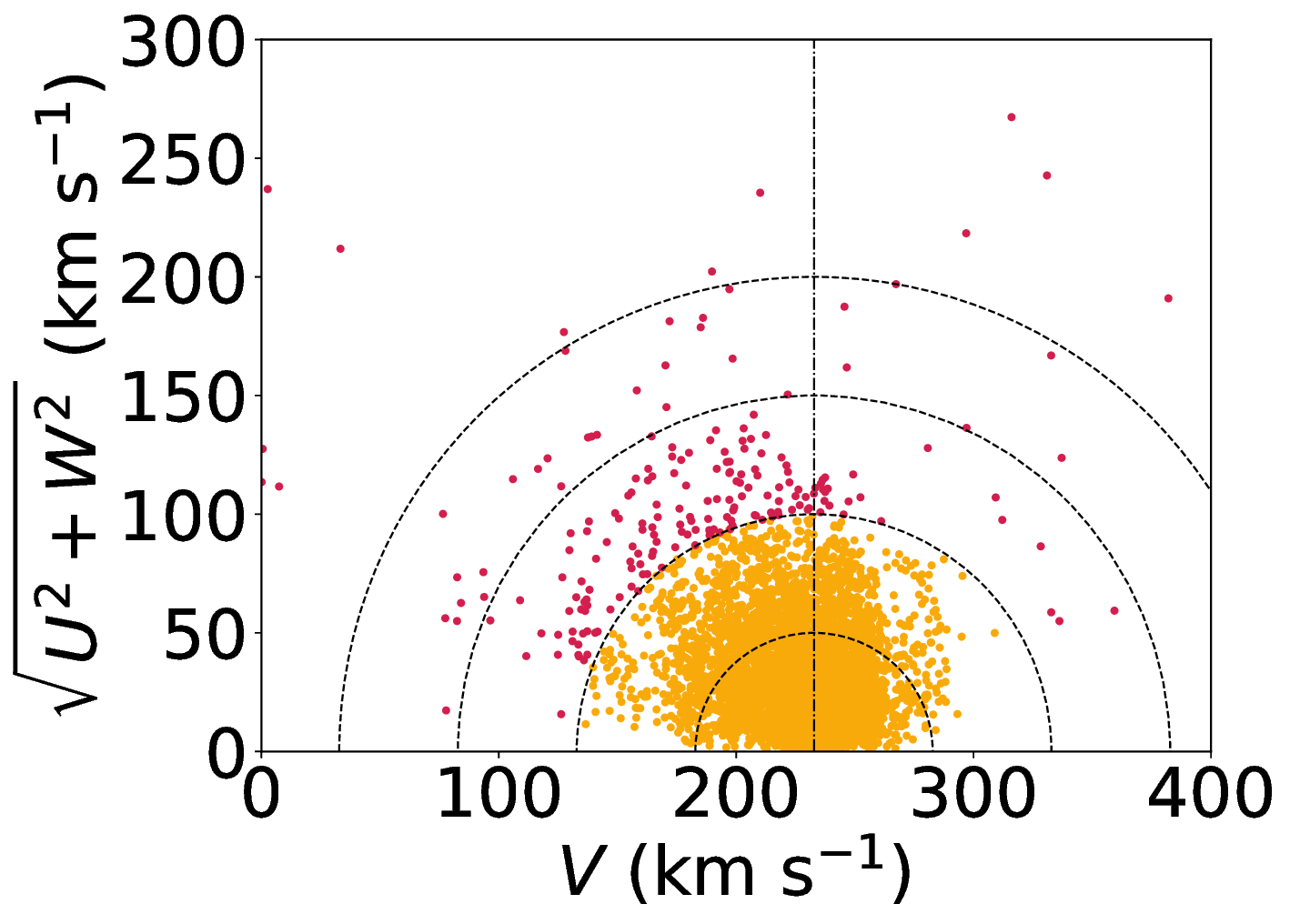} 
    \caption{\small Toomre diagram, with concentric circles centred at 232.8 $\rm{km\,s^{-1}}$ marking velocity thresholds in 50 $\rm{km\,s^{-1}}$ steps. Velocities were corrected using the Solar motion $(U_\odot, V_\odot, W_\odot) = (8.6 \pm 0.9,\, 13.9 \pm 1.0,\, 7.1 \pm 1.0)$ $\rm{km\,s^{-1}}$ \citep{McMillan2017}. Stars exceeding 100 $\rm{km\,s^{-1}}$ relative to the Sun are likely local halo members (228 stars), while the majority correspond to the Galactic disc (a total of 5970 stars). A small fraction (24 corresponding to 0.39\% of the solar total sample) of high-velocity stars were omitted to enhance clarity.
    }
    \label{fig:kinematics}
\end{figure}

\begin{figure}[!ht]
    \centering
    \includegraphics[width=0.7\linewidth]{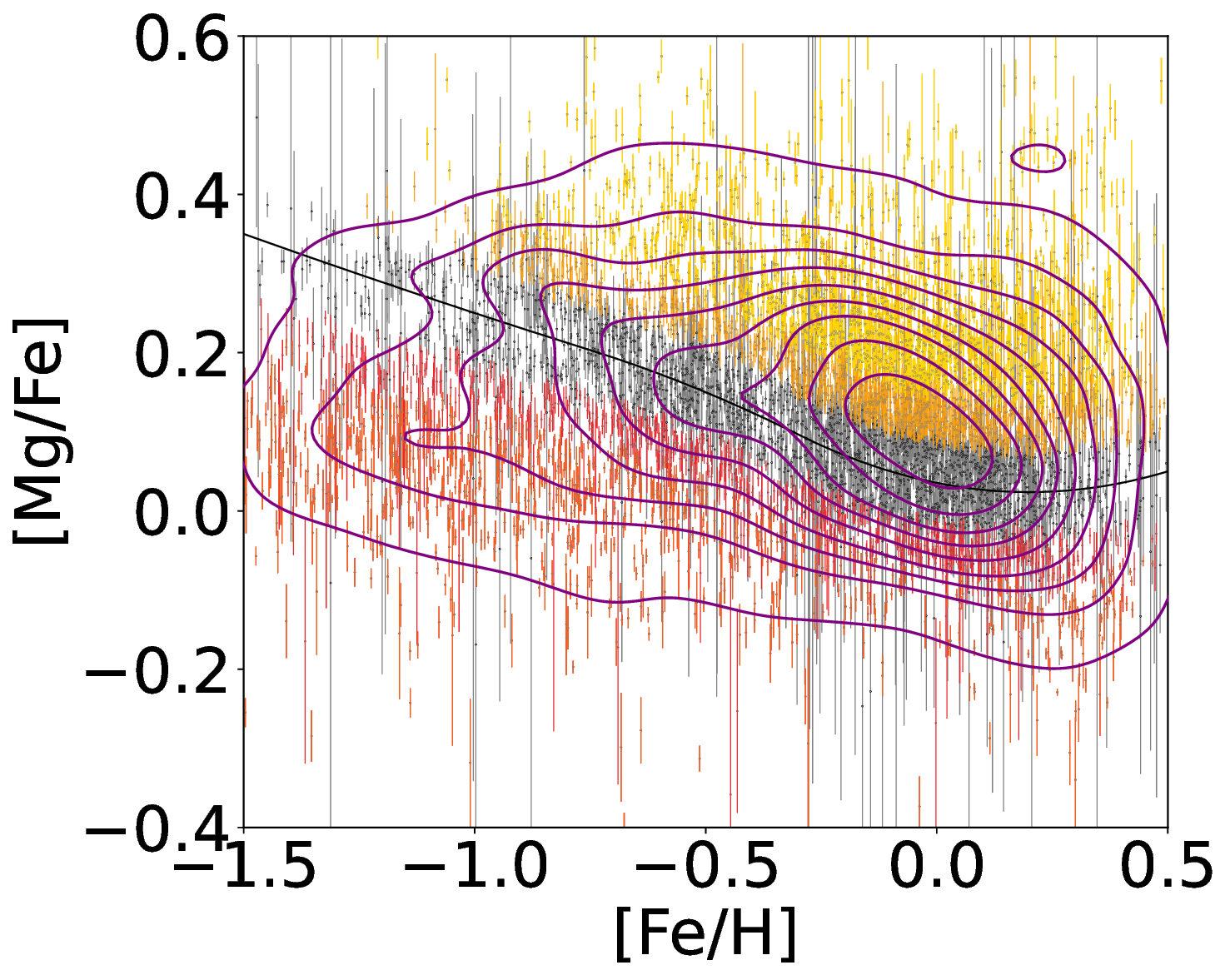}
    \caption{\small Tinsley--Wallerstein diagram, showing the separation between thin and thick disc populations following \citet{RecioBlanco2014}, with the boundary smoothed using a quadratic spline as described in \citet{Dantas2025a}. Grey points lie within 1$\sigma$ of the spline boundary. Stars most likely belonging to the thin disc are shown in shades of red and those more likely belonging to the thick disc are in yellow, with opacity reflecting classification confidence (1–2$\sigma$).
    }
    \label{fig:tw}
\end{figure}

The Toomre diagram (Fig. \ref{fig:kinematics}) reveals a predominantly disc population, with only a small fraction of halo stars, as expected for the solar neighborhood. A small number of 24 (0.39\%) stars were excluded to enhance the clarity and readability of the diagram. To separate thin and thick disc members, we employed the Tinsley–Wallerstein diagram \citep[Fig. \ref{fig:tw};][]{Wallerstein1962, Tinsley1979}, following the criteria of \citet{RecioBlanco2014} with a slight modification as described in \citet[][where a quadratic spline was applied to smooth the boundary]{Dantas2025a}. Among the thick disc stars, we find that the metal-rich, $\alpha$-enhanced component dominates, which is likely reflecting the combined selection effects of the GCNS and GALAH DR4 catalogues.

\vspace{-0.5cm}

The next stages of our study include a more detailed characterization of the sample, with emphasis on the chemical abundances, dynamical properties, and individual spectra of each star.

\begin{acknowledgements}
The authors acknowledge CNPq Project 129985/2025-2 and IFF. PHRA also acknowledges colleagues who provided discussion and technical assistance; MLLD acknowledges the support of Agencia Nacional de  Investigación y Desarrollo (ANID), Chile, through Fondecyt Postdoctorado Folio 3240344. MLLD also aknowledges ANID Basal Project FB210003.
\end{acknowledgements} 
\printbibliography

@ARTICLE{Marigo2017,
       author = {{Marigo}, Paola and others},
        %title = "{A New Generation of PARSEC-COLIBRI Stellar Isochrones Including the TP-AGB Phase}",
      journal = {\apj},
     keywords = {stars: abundances, stars: AGB and post-AGB, stars: carbon, stars: evolution, stars: general, stars: mass loss, Astrophysics - Solar and Stellar Astrophysics},
         year = 2017,
        month = jan,
       volume = {835},
       number = {1},
          eid = {77},
        pages = {77},
          doi = {10.3847/1538-4357/835/1/77},
archivePrefix = {arXiv},
       eprint = {1701.08510},
 primaryClass = {astro-ph.SR},
       adsurl = {https://ui.adsabs.harvard.edu/abs/2017ApJ...835...77M},
      adsnote = {Provided by the SAO/NASA Astrophysics Data System}
}

@ARTICLE{Bressan2012,
    author = {{Bressan}, Alessandro and others},
    %title = "{PARSEC: stellar tracks and isochrones with the PAdova and TRieste Stellar Evolution Code}",
    journal = {\mnras},
    keywords = {stars: evolution, Hertzsprung{\ensuremath{-}}Russell and colour magnitude diagrams, stars: interiors, stars: low-mass, Astrophysics - Solar and Stellar Astrophysics},
    year = 2012,
    month = nov,
    volume = {427},
    number = {1},
    pages = {127-145},
    doi = {10.1111/j.1365-2966.2012.21948.x}
}

@ARTICLE{Bovy2015,
    author = {{Bovy}, Jo},
    %title = "{galpy: A python Library for Galactic Dynamics}",
    journal = {\apjs},
    keywords = {galaxies: general, galaxies: kinematics and dynamics, Galaxy: fundamental parameters, Astrophysics - Astrophysics of Galaxies, Astrophysics - Instrumentation and Methods for Astrophysics},
    year = 2015,
    month = feb,
    volume = {216},
    number = {2},
    %eid = {29},
    pages = {29},
    doi = {10.1088/0067-0049/216/2/29}
}

@ARTICLE{Buder2025,
    author = {{Buder}, Sven and others},
    %title = "{The GALAH survey: Data release 4}",
    journal = {\pasa},
    keywords = {Surveys, the Galaxy, methods: observational, methods: data analysis, stars: fundamental parameters, stars: abundances, Astrophysics - Astrophysics of Galaxies, Astrophysics - Solar and Stellar Astrophysics},
    year = 2025,
    month = may,
    volume = {42},
    %eid = {e051},
    pages = {e051},
    doi = {10.1017/pasa.2025.26},
    archivePrefix = {arXiv},
    eprint = {2409.19858},
    primaryClass = {astro-ph.GA},
    adsurl = {https://ui.adsabs.harvard.edu/abs/2025PASA...42...51B},
    adsnote = {Provided by the SAO/NASA Astrophysics Data System}
}

@ARTICLE{Cui2012,
    author = {{Cui}, Xiang-Qun and others},
    %title = "{The Large Sky Area Multi-Object Fiber Spectroscopic Telescope (LAMOST)}",
    journal = {Research in Astronomy and Astrophysics},
    year = 2012,
    month = sep,
    volume = {12},
    number = {9},
    pages = {1197-1242},
    doi = {10.1088/1674-4527/12/9/003},
    adsurl = {https://ui.adsabs.harvard.edu/abs/2012RAA....12.1197C},
    adsnote = {Provided by the SAO/NASA Astrophysics Data System}
}

@ARTICLE{Dantas2025a,
    author = {{Dantas}, M.~L.~L. and others},
    %title = "{Probing the origins: I. Generalised additive model inference of birth radii for Milky Way stars in the solar vicinity}",
    journal = {\aap},
    keywords = {methods: statistical, stars: abundances, Galaxy: abundances, Galaxy: evolution, Galaxy: stellar content, Galaxy: kinematics and dynamics, Astrophysics of Galaxies, Earth and Planetary Astrophysics, Instrumentation and Methods for Astrophysics, Solar and Stellar Astrophysics},
    year = 2025,
    month = apr,
    volume = {696},
    %eid = {A205},
    pages = {A205},
    doi = {10.1051/0004-6361/202453034},
    archivePrefix = {arXiv},
    eprint = {2502.20441},
    primaryClass = {astro-ph.GA},
    adsurl = {https://ui.adsabs.harvard.edu/abs/2025A&A...696A.205D},
    adsnote = {Provided by the SAO/NASA Astrophysics Data System}
}

@ARTICLE{Gaia2016,
    author = {{Gaia Collaboration} and {Prusti}, T. and others},
    %title = "{The Gaia mission}",
    journal = {\aap},
    keywords = {space vehicles: instruments, Galaxy: structure, astrometry, parallaxes, proper motions, telescopes, Astrophysics - Instrumentation and Methods for Astrophysics},
    year = 2016,
    month = nov,
    volume = {595},
    %eid = {A1},
    pages = {A1},
    doi = {10.1051/0004-6361/201629272}
}

@ARTICLE{GaiaGCNS,
    author = {{Gaia Collaboration} and {Smart}, R.~L. and others},
    %title = "{Gaia Early Data Release 3. The Gaia Catalogue of Nearby Stars}",
    journal = {\aap},
    keywords = {catalogs, astrometry, stars: luminosity function, mass function, Hertzsprung-Russell and C-M diagrams, stars: low-mass, solar neighborhood, Astrophysics - Solar and Stellar Astrophysics, Astrophysics - Astrophysics of Galaxies},
    year = 2021,
    month = may,
    volume = {649},
    %eid = {A6},
    pages = {A6},
    doi = {10.1051/0004-6361/202039498},
    archivePrefix = {arXiv},
    eprint = {2012.02061},
    primaryClass = {astro-ph.SR},
    adsurl = {https://ui.adsabs.harvard.edu/abs/2021A&A...649A...6G},
    adsnote = {Provided by the SAO/NASA Astrophysics Data System}
}

@ARTICLE{Gilmore2022,
    author = {{Gilmore}, G. and others},
    %title = "{The Gaia-ESO Public Spectroscopic Survey: Motivation, implementation, GIRAFFE data processing, analysis, and final data products}",
    journal = {\aap},
    keywords = {Galaxy: stellar content, Galaxy: kinematics and dynamics, stars: abundances, methods: observational, techniques: spectroscopic, surveys, Astrophysics - Solar and Stellar Astrophysics, Astrophysics - Earth and Planetary Astrophysics, Astrophysics - Astrophysics of Galaxies, Astrophysics - Instrumentation and Methods for Astrophysics},
    year = 2022,
    month = oct,
    volume = {666},
    %eid = {A120},
    pages = {A120},
    doi = {10.1051/0004-6361/202243134}
}

@ARTICLE{Majewski2017,
    author = {{Majewski}, Steven R. and others},
    %title = "{The Apache Point Observatory Galactic Evolution Experiment (APOGEE)}",
    journal = {\aj},
    keywords = {Galaxy: abundances, Galaxy: evolution, Galaxy: formation, Galaxy: kinematics and dynamics, Galaxy: stellar content, Galaxy: structure, Astrophysics - Instrumentation and Methods for Astrophysics, Astrophysics - Astrophysics of Galaxies},
    year = 2017,
    month = sep,
    volume = {154},
    number = {3},
    %eid = {94},
    pages = {94},
    doi = {10.3847/1538-3881/aa784d},
    archivePrefix = {arXiv},
    eprint = {1509.05420},
    primaryClass = {astro-ph.IM},
    adsurl = {https://ui.adsabs.harvard.edu/abs/2017AJ....154...94M},
    adsnote = {Provided by the SAO/NASA Astrophysics Data System}
}

@ARTICLE{McMillan2017,
    author = {{McMillan}, Paul J.},
    %title = "{The mass distribution and gravitational potential of the Milky Way}",
    journal = {\mnras},
    keywords = {methods: statistical, Galaxy: fundamental parameters, Galaxy: kinematics and dynamics, Galaxy: structure, Astrophysics - Astrophysics of Galaxies},
    year = 2017,
    month = feb,
    volume = {465},
    number = {1},
    pages = {76-94},
    doi = {10.1093/mnras/stw2759}
}

@ARTICLE{Randich2022,
    author = {{Randich}, S. and others},
    % title = "{The Gaia-ESO Public Spectroscopic Survey: Implementation, data products, open cluster survey, science, and legacy}",
    journal = {\aap},
    keywords = {surveys, catalogs, techniques: spectroscopic, stars: fundamental parameters, stars: abundances, open clusters and associations: general, Astrophysics - Astrophysics of Galaxies, Astrophysics - Solar and Stellar Astrophysics},
    year = 2022,
    month = oct,
    volume = {666},
    %eid = {A121},
    pages = {A121},
    doi = {10.1051/0004-6361/202243141}
}

@ARTICLE{RecioBlanco2014,
    author = {{Recio-Blanco}, A. and others},
    %title = "{The Gaia-ESO Survey: the Galactic thick to thin disc transition}",
    journal = {\aap},
    keywords = {Galaxy: abundances, Galaxy: disk, Galaxy: stellar content, stars: abundances, Astrophysics - Astrophysics of Galaxies},
    year = 2014,
    month = jul,
    volume = {567},
    %eid = {A5},
    pages = {A5},
    doi = {10.1051/0004-6361/201322944}
}

@ARTICLE{Tinsley1979,
    author = {{Tinsley}, B.~M.},
    %title = "{Stellar lifetimes and abundance ratios in chemical evolution.}",
    journal = {\apj},
    keywords = {Abundance, Chemical Evolution, Galactic Evolution, Life (Durability), Nuclear Fusion, Stellar Evolution, Carbon, Iron, Mass Ratios, Oxygen, Stellar Mass, Supernovae, Time Dependence, White Dwarf Stars, Astrophysics, Nucleosynthesis:Stellar Evolution},
    year = 1979,
    month = may,
    volume = {229},
    pages = {1046-1056},
    doi = {10.1086/157039},
    adsurl = {https://ui.adsabs.harvard.edu/abs/1979ApJ...229.1046T},
    adsnote = {Provided by the SAO/NASA Astrophysics Data System}
}

@ARTICLE{Xiang2019,
    author = {{Xiang}, Maosheng and others},
    %title = "{Abundance Estimates for 16 Elements in 6 Million Stars from LAMOST DR5 Low-Resolution Spectra}",
    journal = {\apjs},
    keywords = {Spectroscopy, Spectroscopic binary stars, Stellar atmospheres, Stellar abundances, Astronomy databases, Stellar properties, Stellar spectral lines, Astronomy data analysis, Sky surveys, Fundamental parameters of stars, Milky Way Galaxy, Astronomy data modeling, 1558, 1557, 1584, 1577, 83, 1624, 1630, 1858, 1464, 555, 1054, 1859, Astrophysics - Solar and Stellar Astrophysics, Astrophysics - Astrophysics of Galaxies, Astrophysics - Instrumentation and Methods for Astrophysics},
    year = 2019,
    month = dec,
    volume = {245},
    number = {2},
    %eid = {34},
    pages = {34},
    doi = {10.3847/1538-4365/ab5364},
    archivePrefix = {arXiv},
    eprint = {1908.09727},
    primaryClass = {astro-ph.SR},
    adsurl = {https://ui.adsabs.harvard.edu/abs/2019ApJS..245...34X},
    adsnote = {Provided by the SAO/NASA Astrophysics Data System}
}

@ARTICLE{Wallerstein1962,
    author = {{Wallerstein}, George},
    %title = "{Abundances in G. Dwarfs.VI. a Survey of Field Stars.}",
    journal = {\apjs},
    year = 1962,
    month = feb,
    volume = {6},
    pages = {407},
    doi = {10.1086/190067},
    adsurl = {https://ui.adsabs.harvard.edu/abs/1962ApJS....6..407W},
    adsnote = {Provided by the SAO/NASA Astrophysics Data System}
}
\end{document}